\documentclass{article}
\usepackage[dvips]{graphicx}  %  XXX archive  version

\begin{document}

\newcommand{\rum}{\rule{0.5pt}{0pt}}

\newcommand{\rub}{\rule{1pt}{0pt}}

\newcommand{\rim}{\rule{0.3pt}{0pt}}

\newcommand{\numtimes}{\mbox{\raisebox{1.5pt}{${\scriptscriptstyle \times}$}}}

\renewcommand{\refname}{References}

%\twocolumn[%

\begin{center}

{\Large\bf  Re-Analysis of the Marinov Light-Speed Anisotropy Experiment 
\rule{0pt}{13pt}}\par

\bigskip

Reginald T. Cahill \\ 

{\small\it School of Chemistry, Physics and Earth Sciences, Flinders University,

Adelaide 5001, Australia\rule{0pt}{13pt}}\\

\raisebox{-1pt}{\footnotesize E-mail: Reg.Cahill@flinders.edu.au}\par

\bigskip\smallskip

{\small\parbox{11cm}{%

The anisotropy of the speed of light at 1 part in $10^3$ has been detected by Michelson and Morley (1887), Miller (1925/26), Illingworth (1927), Joos (1930), Jaseja {\it et al.} (1964), Torr and Kolen (1984), De Witte (1991) and Cahill (2006) using a variety  of experimental techniques, from gas-mode Michelson interferometers (with the relativistic theory for these only determined in 2002)  to one-way RF coaxial cable propagation timing. All agree on the speed, right ascension and declination of the anisotropy velocity.  The Stephan Marinov experiment (1984) detected a light speed anisotropy using a mechanical coupled shutters technique which has holes in co-rotating disks, essentially a one-way version of the Fizeau mechanical round-trip speed-of-light experiment.  The Marinov data is re-analysed  herein because the velocity vector he determined is in a very different direction to that from the above experiments. No explanation for this difference has been uncovered. 
\rule[0pt]{0pt}{0pt}}}\bigskip

\end{center}
%]

\section{Introduction}

That the speed of light in vacuum is the same in all directions, i.e. isotropic,  for all observers has been taken as  a critical assumption  in  the standard formulation of fundamental physics, and was introduced by Einstein in 1905 as one of his key postulates when formulating his interpretation of Special Relativity.
The need to detect any anisotropy has challenged physicists from the 19th century to the present day, particularly  following the Michelson-Morley experiment of 1887. 
The problem arose when Maxwell in 1861 successfully computed the speed of light  $c$ from his unified theory of electric and magnetic fields: but what was the speed $c$ relative to?  There have been many attempts to detect any supposed light-speed anisotropy, and as discussed in the  Sect.2 there have so far been 8 successful  and consistent such experiments, and as well numerous unsuccessful experiments, i.e. experiments in which no anisotropy was observed. The reasons for these different outcomes is now understood:  any light-speed anisotropy produces not only an expected `direct' effect, being that which is expected to produce a `signal', but also affects  the very physical structure of the apparatus, and with this effect usually overlooked in the design of some detectors.  In some designs these effects exactly cancel. 

The key point here is not whether the predicted   Special Relativity effects are valid or invalid, for the experimental evidence is overwhelming that these predictions are valid, but rather whether the Lorentz or Einstein {\it interpretation}  of Special Relativity is correct.     This debate has always been confused by the failure to understand that the successes of Special Relativity, and its apparent deduction from the above Einstein postulate, does not actually require that the speed of light be invariant, as Fitzgerald and Lorentz pointed out over 100 years ago, see discussions in \cite{Book,Coax}. Rather the issue is whether the Special Relativity effects are caused by absolute motion of systems through a dynamical 3-space, or whether we have no 3-space and only a four-dimensional spacetime. So the question  is about whether or not  the 3-space can be detected by means of the anisotropy of light, since in this interpretation the speed is $c$ only relative to this space locally. This comes down to the issue of whether 3-space or spacetime actually exists, not whether the local Special Relativity effects are valid or not. 

As already stated there is overwhelming evidence  from 8 experiments that the speed of light is anisotropic, and with a {\it large} anisotropy at the level of 1 part in $10^3$: so these experiments show that a dynamical 3-space exists, and that the spacetime concept was only a mathematical construct - it does not exist as an entity of reality, it has no ontological significance.   These developments have lead to a new physics in which the dynamics of the 3-space have been formulated, together with the required generalisations of the Maxwell  equations (as first suggested by Hertz in  1890 \cite{Hertz}), and of  the Schr\"{o}dinger and Dirac equations, which have lead to the new emergent theory and explanation of gravity, with numerous confirmations of that theory from the data from black hole systematics, light bending, spiral galaxy rotation anomalies, bore hole anomalies, etc. This data has revealed that the coupling constant for the self-interaction of the dynamical 3-space is none other than the fine structure constant  $\approx 1/137$ \cite{alpha,DM,galaxies,boreholes}, which suggests an emerging unified theory of quantum matter and a quantum foam description of the dynamical 3-space.   

The substance of this report is to re-examine a light-speed anisotropy experiment performed by Stefan Marinov in Graz in 1984  using essentially separated mechanical light choppers   in a special modified version of the original Fizeau technique - basically Marinov  measured the {\it difference}   in travel time between light beams travelling in opposite directions \cite{Marinov}.  However there is an apparent problem with the anisotropy velocity vector that Marinov reported, namely it had a very different direction, and also a somewhat different speed, from that which has recently been determined from the 8 experiments discussed in Sect.2.    
However we conclude that no explanation has been found for why the Marinov velocity vector is so different from that determined by the 8 other experiments.

 It is truly amazing that for over 100 year physics has  failed to acknowledge the anisotropy of the light speed; and a very large effect of the order of 1 part in $10^3$.  This is partly explained by the fact that any discussion of these experiments and the implications is banned from mainstream physics.  This apparently follows from the long-standing misconception that the successes of Special Relativity, and of Lorentz symmetry, require that no such anisotropy exists.  Rather, the existence of a preferred direction, of an actual locally detected  frame of reference, is perfectly consistent with Special Relativity and Local Lorentz Symmetry, although the explication of this is somewhat subtle, requiring a very careful  operational definition of what is meant by the space and time coordinates in the different formalisms.  Essentially the well-known Einstein formalism builds into the definitions of space and time coordinates that the speed of light is invariant. However  such definitions, while permitted  mathematically, do not correspond to the physical definition.

\section{Brief History of Light Speed Anisotropy Measurements\label{section:history}}
The most famous and influential of the early attempts to detect any anisotropy in the speed of light was the Michelson-Morley experiment of 1887, \cite{MM}. Despite that, and its influence on physics, its operation was only finally understood in 2002 \cite{MMCK,AMGE,MMC}.   The problem has been that the Michelson interferometer has a major flaw in its design, when used to detect any light-speed anisotropy effect\footnote{Which also severely diminishes its use in long-baseline interferometers built to detect gravitational waves}.  To see this requires use of Special Relativity effects.   The Michelson interferometer compares the round-trip light travel time in two orthogonal arms, by means of interference fringe shifts measuring time differences, as the device is rotated. However  if the device is operated in vacuum, any anticipated change in  the total travel times caused by the light travelling at different speeds in the outward and inward directions is exactly cancelled by the Fitzgerald-Lorentz mirror-supporting-arm contraction effect - a real physical effect. Of course this is precisely how Fitzgerald and Lorentz independently arrived at the idea of the length contraction effect.  In vacuum this means that the round-trip travel times in each arm {\it do not} change during rotation. This is the fatal  design flaw that has confounded physics for over 100 years.  However the cancellation  of a supposed change in the round-trip travel times and the Lorentz contraction effect is merely an incidental flaw of the Michelson interferometer. The critical observation  is that if we have a gas in the light path, the round-trip travel times are changed, but the Lorentz arm-length contraction effect is unchanged, and then these effects no longer exactly cancel. Not surprisingly the fringe shifts are now  proportional to $n^2-1$, where $n$ is the refractive index of the gas.   Of course with  a gas present one must also take account of the Fresnel drag effect, because the gas itself is in absolute motion. This is an important effect, so large  in fact that it reverses the sign of the time differences between the two arms, although in operation that is not a problem.  As well, since for example for air $n=1.00029$ at STP, the sensitivity of the interferometer is very low. Nevertheless the Michelson-Morley experiment as well as the Miller Michelson interferometer experiment of 1925/1926 \cite{Miller} were done in air, which is why they indeed observed and reported  fringe shifts. As well  Illingworth \cite{C5} and Joos \cite{C6} used helium gas in the light paths in their Michelson interferometers; taking account of that brings their results into agreement with those of the air interferometer experiment, and so confirming the refractive index effect.   Jaseja {\it et al.} \cite{C7} used a He-Ne gas mixture of unknown refractive index, but again detected fringe shifts on rotation.  A re-analysis of the data from the above experiments, particularly from the enormous data set of Miller,  has revealed that a large light-speed  anisotropy had been detected from the very beginning of such experiments, where the speed is some $430\pm 20$km/s - this is in excess of 1 part in $10^3$, and the Right Ascension and Declination of the direction was determined by Miller \cite{Miller} long ago.  

Curiously numerous experimentalists  developed vacuum mode Michelson interferometers as vacuum pump technology became available, and of  course the fringe shifts eventually went away, supposedly confirming that no light-speed anisotropy existed.   However one must always be careful of so-called ``null'' experiments - it may actually be a ``dud'' experiment instead. In recent years the vacuum-mode interferometers have been `improved' considerably by using small cryogenic vacuum-mode Fabry-Perot resonators, as for example M\"{u}ller {\it et al.} \cite{FP}.  Trying to get experimentalists to put some gas in at least one of the resonators, so that the gas effect enables the device to detect the anisotropy, has proven to be very challenging\footnote{S. Dawkins and A. Luiten from the University of Western Australia have now done just that, putting $N_2$ gas  into one arm. At the Australian Institute of Physics 17th Congress in Brisbane, Australia, in December 2006, they reported that beat frequency shifts, the analogue of fringe shifts,  were now detected as the earth rotated - because of extremely good stability they don't need to do short term rotations of the apparatus.  Of course we must wait until they have optimised the apparatus and  reported their results }. 

Another technique that has been successfully used is to measure the one-way travel time of RF waves in  coaxial cables, as in Torr and Kolen 1981 \cite{Torr} with the one-way travel through 500m of cable, De Witte 1991 \cite{DeWitte} using travel time differences between two 1.5km cables, and Cahill 2006 \cite{Coax} using    two 5 meter cables facilitated by the optical fiber effect for orientation-invariant timing transfers.  Over the years the problem of making very accurate timing measurements that are stable over days has evolved. Torr and Kolen and De Witte both used  multiple atomic clocks, and long coaxial cables, while Cahill  uses one atomic clock and the optical fiber effect. These experiments are discussed in \cite{Coax}.  In the DeWitte and Cahill  experiments one measures the {\it difference}  in travel time between RF waves travelling in opposite directions. The results from these 3 coaxial cable experiments and the earlier gas-mode Michelson interferometer experiments are in excellent agreement.  Also as discussed in \cite{Coax} the optical fiber effect permits the construction of very small 1st order in v/c interferometers without the gas $n^2-1$ effect, and these are extremely accurate and cheap.   These differential one-way coaxial-cable time-difference experiments are  analogous to the Marinov mechanical Coupled Shutter device, which we now finally discuss.

\begin{figure}[t]
\vspace{0mm}
\hspace{30mm}
\setlength{\unitlength}{1.5mm}
\hspace{30mm}\begin{picture}(0,20)
\thicklines
\put(30,-19){{\bf $S$}}\put(36,-14){{\bf $BS$}}\put(29,-6){{\bf $D$}}\put(-7.5,4){{\bf $D$}}
\put(34.5,9.5){{\bf $M$}}

\put(7,-0.2){\line(1,0){16}}
\put(7,-0.3){\line(1,0){16}}
\put(7,-0.4){\line(1,0){16}}

\put(-6,9){\line(1,0){9}}\put(5.5,9){\line(1,0){18}}\put(25.5,9){\line(1,0){8}}
\put(14,9){\vector(-1,0){1}}\put(-2,9){\vector(-1,0){1}}\put(29,9){\vector(-1,0){1}}
\put(14,-9){\vector(1,0){1}}\put(-2,-9){\vector(1,0){1}}\put(27,-9){\vector(1,0){1}}
\put(-6,-13){\line(1,0){39.5}}\put(15,-13){\vector(-1,0){1}}
\put(33.5,-13){\line(0,1){22.0}}\put(33.5,0){\vector(0,1){1}}

\put(-6,-9){\line(1,0){9}}\put(5.5,-9){\line(1,0){18}}\put(25.5,-9){\line(1,0){4}}
\put(29.5,-11){\line(0,1){4}} \put(30.5,-11){\line(0,1){4}} \put(29.5,-11){\line(1,0){1}}
\put(29.5,-7){\line(1,0){1}}

\put(32.2,10.5){\line(1,-1){3}}\put(-7.2,-11.5){\line(1,-1){3}}\put(-7,-10){\line(1,1){3}}
\put(-6,-13){\line(0,1){4}}

\put(-7,7){\line(0,1){4}} \put(-6,7){\line(0,1){4}} \put(-7,7){\line(1,0){1}}
\put(-7,11){\line(1,0){1}}

\put(32.2,-11.5){\line(1,-1){3}}\put(32.0,-14.5){\line(1,0){3}}\put(32.2,-14.5){\line(0,1){3}}
\put(33.5,-17.5){\line(0,1){6}}\put(33.5,-15.5){\vector(0,1){1}}

\put(32.5,-20){\line(0,1){3}} \put(34.5,-20){\line(0,1){3}} \put(32.5,-17){\line(1,0){2}}
\put(32.5,-20){\line(1,0){2}}

\put(3,10){\line(1,0){1.5}}

\put(3,-10){\line(1,0){1.5}}
\bezier{500}(0,0)(0,10)(3,10)
\bezier{500}(0,0)(0,-10)(3,-10)
\bezier{500}(6,0)(6,10)(3,10)
\bezier{500}(6,0)(6,-10)(3,-10)

\bezier{500}(7,0)(7,10)(4,10)
\bezier{500}(7,0)(7,-10)(4,-10)

\put(3,9){\circle*{1}}
\put(3,-9){\circle*{1}}
\put(1.0,0.2){\circle*{1}}
\put(1.3,4.5){\circle*{1}}\put(4.6,4.2){\circle*{1}}
\put(1.3,-4.5){\circle*{1}}\put(4.6,-4.9){\circle*{1}}
\put(5,-0.2){\circle*{1}}

\put(23,10){\line(1,0){1.5}}
\put(23,-10){\line(1,0){1.5}}
\bezier{500}(20,0)(20,10)(23,10)
\bezier{500}(20,0)(20,-10)(23,-10)
\bezier{500}(26,0)(26,10)(23,10)
\bezier{500}(26,0)(26,-10)(23,-10)

\bezier{500}(27,0)(27,10)(24,10)
\bezier{500}(27,0)(27,-10)(24,-10)

\put(23,9){\circle*{1}}
\put(23,-9){\circle*{1}}
\put(21,0.2){\circle*{1}}
\put(21.3,4.5){\circle*{1}}\put(24.6,4.2){\circle*{1}}
\put(21.3,-4.5){\circle*{1}}\put(24.6,-4.9){\circle*{1}}
\put(25,-0.2){\circle*{1}}

\end{picture}
\vspace{30mm}\caption{\small{Schematic diagram of the Marinov one-way speed-of-light apparatus using the  ``coupled shutters'' technique.  Two co-rotating disks have holes through which light passes.  $S$ is the light source, $BS$ is a beam splitter, and various mirrors $M$ direct the light though the holes. The light intensity is measured by the photocell detectors $D$. Changes in the speed of light affect the amount of light that can pass through the distant hole.}}
\vspace{5mm}

\end{figure}
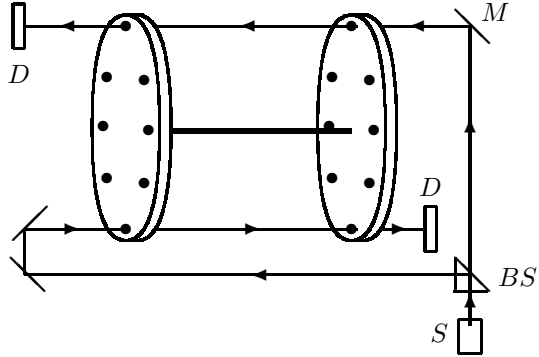

\begin{figure}[t]
\hspace{30mm}\includegraphics[scale=0.44]{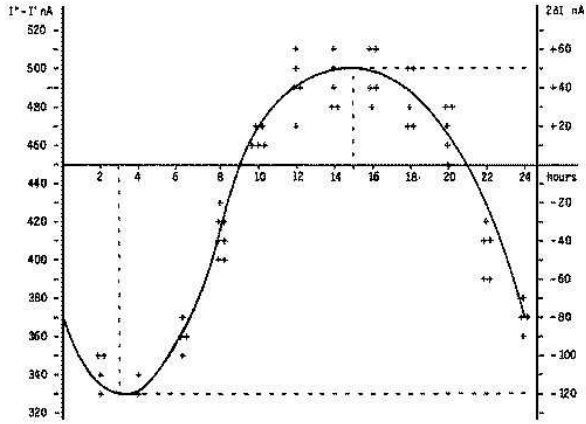}
\vspace{-20mm}\caption{\small{ Measurements of photocell current differences, 2$\delta I$, from the Marinov ``coupled-shutters'' one-way light-speed experiment in Graz, Austria,  February 9-13, 1984, reproduced from \cite{Marinov}.  The times in hours are local times.   The ``null line'' (i.e the abscissa) turns out to be arbitrary, as Marinov did not establish the value of the asymmetry  speed $V$ in (\ref{eqn:V}) of the  detector, and in fact incorrectly assumed that $V=0$.   }
\label{fig:Fig1}}\end{figure}

\begin{figure}[t]
\hspace{40mm}\includegraphics[scale=0.6]{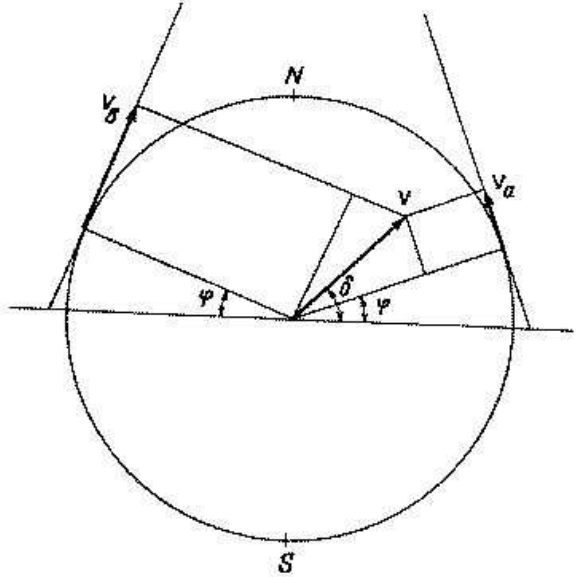}
\vspace{-25mm}\caption{\small{  Section of the Earth showing the NS direction,  reproduced from \cite{Marinov}. The absolute velocity $\bf v$ of the flow of space past the earth, has declination $\delta$, and Graz has Latitude $\phi=45$ degrees N. The speeds $v_a$ and $v_b$ are the projections  of $\bf v$ onto the  detector,  lying along the  local NS meridian,  at the extremes, leading to the minimum and maximum shown in Fig.1.  Expressions for $v_a$ and $v_b$ are given in (\ref{eqn:Va}), but taking into account a significant asymmetry effect, equivalent  to adding the speed  $V$, we must use the expressions in (\ref{eqn:V}). We report the final result using the Miller convention that we specify the speed and direction of the Earth through the local space, i.e the right ascension and declination of the  velocity $-\bf v$, see Fig.3.}
\label{fig:Fig2}}\end{figure}

\begin{figure}[t]

\hspace{35mm}\includegraphics[scale=1.3]{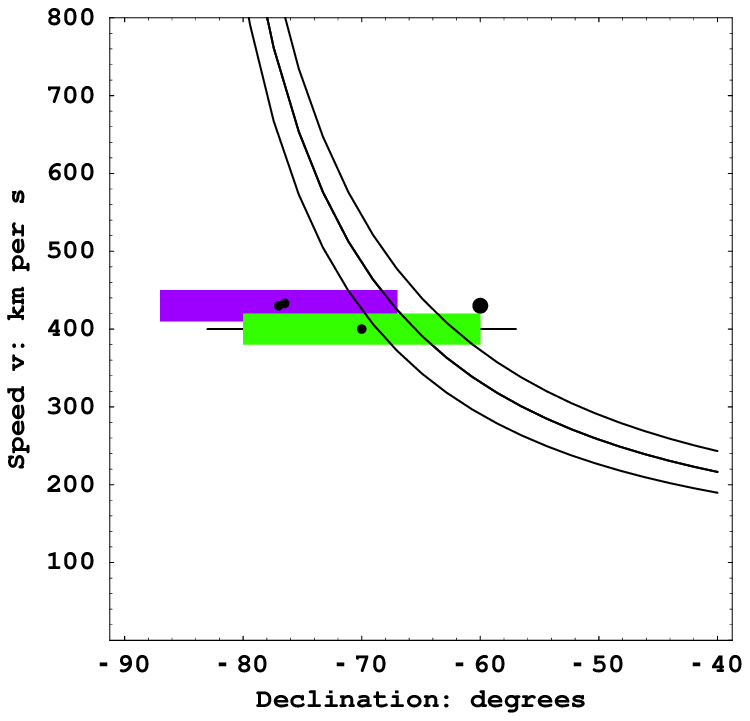}
\caption{\small{ The curved lines show the Marinov speed-declination (v-$\delta$) plots from (\ref{eqn:vdelta}) for February 8-13, 1984, with the outer lines defining the range that follows from Marinov's stated uncertainty in $v_a$ and $v_b$ of $\pm15$km/s, but with the speed $v$ scaled down by a factor of 2.  The upper-left box, indicating uncertainties and/or wave fluctuations, with center indicated by left-most dot, is the speed and declination in February 1926 from the Miller  gas-mode Michelson interferometer \cite{Book,Miller}. The right-most dot in this Miller box is the result, believed to be from February 1981, without computed uncertainties,  from the Torr-Kolen coaxial cable experiment   \cite{Torr},  see \cite{Book,Coax} for details, and which is in remarkable agreement with the Miller results. The other three results are not from February: the lower-right box with center indicated by a dot, is from the Cahill coaxial cable experiment of August/September of 2006 \cite{Coax},  the partially obscured horizontal line is the lower-limit speed from the  Michelson-Morley gas-mode interferometer experiment on July 11, 1887 at 7:00 hr local sidereal time \cite{MM}, see \cite{Coax} for re-analysis, and the right-most dot is the data from the coaxial cable 1991 experiment by De Witte with no uncertainties shown  \cite{Book,DeWitte, Coax},   which appears to be from  October.  These results show that the new analysis herein of the Marinov  data reveals that the ``coupled shutters'' technique would agree with the February Miller and Torr-Kolen data if the Marinov speed was scaled down by a factor of 2.  By  assuming  a speed of  some 430 km/s then from the Marinov data we obtain a declination for ${-\bf v}$ of $68\pm 4$ degrees S, although the actual errors are probably larger.  }
\label{fig:Marinov}}\end{figure}

\section{The Marinov Light-Speed Anisotropy Experiment\label{Marinov}}
In 1984 Stephan Marinov performed in Graz a direct measurement of the variations  in the one-way speed of light.  This used the classic rotating shutters method where the speed of light is determined by observing the light transmission intensity when it propagates through vacuum between small holes in separated but co-rotating disks, as shown in Fig.1.  The use of a mechanical timing method to determine the speed of light dates back to Fizeau who in 1849 performed a round-trip speed-of-light measurement in which a beam of light was reflected back  from a mirror 8 km away, with   the beam passing through the gaps between the teeth of a rapidly rotating wheel.   However there is an import development with the Marinov experiment: it is a one-way speed measurement. As discussed above the round-trip time measurements cannot determine the one-way speed of light, except under special circumstance, such as when the light propagates in a gas, as in gas-mode Michelson interferometers and Fabry-Perot resonant cavities.  However the Marinov experiment  ideally measures the time-difference between the travel-time in two opposite directions.  The apparatus involved 2 co-rotating disks separated by a distance of  120 cm. Light from an Argon laser was split and directed axially towards  the disks through small holes at a radius of 12 cm.   Photocells detected the light when it passed through the holes in the opposite disks, with the whole apparatus rotating at 200 rev/s, and with the axis aligned along the local meridian, i.e. NS.  The intensity of the light emerging from the holes in the  disks depends on the light travel time, and is determined  by  means of a galvanometer measuring the current from the silicon photodetectors.  More details are available in \cite{Marinov}. Fig.2 shows the current differences over the 5 days February 9-13.  The data clearly shows the expected time signature.

 The geometry of the experiment is explained in Fig.3.  The key effect is that the speed of light is $c$ relative to the space flowing past the Earth with velocity ${\bf v}$.  This means that the speed of light relative to the axis of the apparatus varies as the Earth rotates, as the angle between the flow and the detector light beams changes.  At the extremes the projected speeds are $v_a$ and $v_b$, and are given by
  \begin{equation}  v_a=v\sin(\delta-\phi), \mbox{ \ \ \   }   v_b=v\sin(\delta+\phi),
\label{eqn:Va} \end{equation}
where $v=|{\bf v}|$.
However there is an important experimental aspect which must be taken into account, namely that the two components of the apparatus, namely that part with the light travelling essentially N to S can never be made identical to the part with the light travelling from S to N at the level of precision required in this experiment.  Marinov acknowledges this problem but in the end actually failed to come to the correct method for dealing with it.  Because of the asymmetry of the two parts of the experiment (\ref{eqn:Va}) must be put in the form
 \begin{equation}  v_a=v\sin(\delta-\phi) +V, \mbox{ \ \ \   }   v_b=v\sin(\delta+\phi) +V
\label{eqn:V} \end{equation} 
where now $v_a$ and $v_b$ are the  speeds determined from the current measurements, and most importantly $V$  is an effective speed that parametrizes the asymmetry in the apparatus: because even if the flow speed $v=0$ the apparatus will register  non-zero $v_a=v_b\neq 0$. The only way to determine $V$ is to rotate the apparatus while the disks are  spinning, from NS to SN orientation.  Then this interchanges the two parts, and now the $v_a$ and $v_b$ are given by (\ref{eqn:V}), but now with $V\rightarrow -V$.  Then one could compute $V$, and then the `zero speed' of the apparatus is properly set.  A similar task arose in the Cahill one-way RF coaxial cable experiment. There timing signals between the ends of the RF cables is facilitated by sending infrared signals through optical fibers for which the propagation times are invariant, unlike the RF in the coaxial cables. The cables cannot be cut to equal length with  sufficient accuracy, and to set the `zero speed' reading of  the device, one could rotate the device through 180 degree, which causes the asymmetry effect to manifest with the opposite sign.    However  in this experiment another solution was available, namely to fold the cables into a circular loop.  Then the  effects of absolute motion cancels, and the `zero' for the instrument is easily established.  While Marinov was certainly aware of this asymmetry problem, in the end he effectively ignored it.  In the re-analysis herein we do take account of this important problem.  It means however that because the two equations in (\ref{eqn:V}) now have  three unknowns, $v, \delta$ and $V$,  we cannot determine a unique solution. Marinov of course assumed that $V=0$, which then permitted a unique but incorrect solution. Eliminating $V$ from (\ref{eqn:V})
we obtain 
   \begin{equation}  v(\delta)=\frac{v_b-v_a}{\sin(\delta+\phi) -\sin(\delta-\phi)},
\label{eqn:vdelta} \end{equation}                  
which at best gives us only a possible relationship between $v$ and $\delta$. In Fig.4 we show various $v-\delta$ results from some five light-speed anisotropy experiments, as explained in the caption. These are remarkably consistent, taking into account that they vary over a year because of the changing velocity of the Earth about the sun.  However the Miller and Torr-Kolen results are from February data, and are thus most relevant to the Marinov experiment.   Marinov reported that  from the data in Fig.2 he deduced that $v_a=-342 \pm 30$ km/s and $v_b=+143\pm 30$ km/s. However the resulting $v-\delta$ plot is then almost exactly twice as large as the $v-\delta$ values from the indicated experiments.  It is possible that there is an error here in going from photodetector currents to the $v_a$, and $v_b$ speeds. So here I have taken $v_a=(-342 \pm 30)/2$ km/s and $v_b=(+143\pm 30)/2$ km/s. These generate the curves in Fig.4.    If we assume the value for $v$ from the Miller and Torr-Kolen experiments of  430 km/s then we obtain a  declination  for $-\bf v$ of $68\pm 4$ degrees S, although the actual errors are probably larger.  For this solution one can now extract the value of $V$ from  (\ref{eqn:V}), and we obtain $V=-228$km/s, confirming that the asymmetry effect is significant.  So one can conclude that, subject to the factor of two correction, the speed and declination from the Marinov data is consistent with the other experiments.   Ignoring the asymmetry speed $V$ Marinov obtained a speed and declination for $\bf v$ of $v=362 \pm 40$km/s, and $\delta=-24\pm 7$ degrees, i.e a declination for  $-\bf v$ of  $\delta=+24\pm 7$ degrees.

Marinov reported a right ascension for $\bf v$ of $\alpha=12.5^h  \pm 1^h$, based upon the local times for the maximum and minimum in Fig.2 of $15^h \pm 1^h$ and  $3^h \pm 1^h$.   Let us  work through this determination. At midday on March 21 the local sidereal time at Greenwich is $0^h$. Graz has longitude $15^0 26'$E, or 1 hour ahead of Greenwich. So  the local  time in Graz of  $13^h$ corresponds to a local sidereal time of $+1^h$. The experiment was done in the period February 9-13, which is approximately 38 days before March 21, and so the local sidereal time was retarded by $2.5^h$, so that on February 11 at $13^h$ in Graz the local sidereal time is $-1.5^h$.  Then the local time of $15^h=13^h+2^h$ corresponds to a local sidereal time of   $\alpha=2^h-1.5^h=0.5^h$, and $3^h=13^h-10^h$ corresponds to   $\alpha=-10^h-1.5^h=-11.5^h \equiv 12.5^h$.  Hence the right ascension of $-\bf v$ from the Marinov experiment is $0.5^h  \pm 1^h$. This is to be compared to the right ascension of  $-\bf v$ reported by Miller for February of  $6^h$. Hence the Marinov data gives a right ascension for ${\bf v}$  of $12.5^h$ which agrees with  that reported by Marinov  \cite{Marinov}.  

\section{The Cosmic Microwave Background  Anisotropy Velocity}
The Cosmic Microwave Background (CMB) velocity  is often confused with the Absolute Motion (AM) velocity or light-speed anisotropy velocity  as determined in the experiments discussed herein. However these are totally unrelated and in fact point in very different directions, being almost at 90$^0$ to each other, with the CMB velocity being 369km/s in direction $(\alpha=11.2^h, \delta=-7.22^0)$.  
The CMB velocity vector was first determined in 1977 by Smoot {\it et al.} \cite{Smoot}  giving 390$\pm$60km/s,  $(\alpha=11\pm0.6^h, \delta=6\pm10^0)$.

The CMB velocity is obtained by defining  a frame  of reference in which the thermalised CMB $3^0$K radiation is isotropic, that is by removing the dipole component, and  the CMB velocity is the velocity of the Earth in that frame.    The CMB velocity is a measure of the motion of the solar system relative to the universe as a whole, or at least a spherical shell of the universe  some 13Gyrs in the past, and indeed the near uniformity of that radiation in all directions demonstrates that we may  meaningfully refer to the spatial structure of the universe.  The concept here is that at the time of decoupling of this radiation from matter that matter was on the whole, apart from small observable fluctuations, on average at rest with respect to the 3-space. So the CMB velocity is  not motion with respect to the  {\it local} 3-space now; that is the AM velocity.   Contributions to the AM  velocity would arise from the orbital motion of the solar system within the Milky Way galaxy, which has  a speed of some 250 km/s, and contributions from the motion of the Milky Way within the local cluster, and so on to perhaps super clusters, as well as flows of space associated with gravity in the Milky Way and local galactic cluster etc.    The difference between the CMB velocity and the AM velocity is explained by the spatial flows that are responsible for  gravity at the galactic scales.

In a recent light-speed anisotropy experiment by Navia {\it et al.} \cite{Navia} it was assumed in the analysis that the light speed anisotropy velocity (AM) is the same as the CMB velocity.

\section{Conclusions}
The re-analysis herein of the Marinov one-way light-speed anisotropy experiment has left unexplained why his anisotropy velocity is so different from that  detected by 8 other experiments. However we note that it is quite similar to the anisotropy vector arising from the CMB detections.  The observed light-speed anisotropy in all the experiments is very large being in excess of 1 part in $10^3$.  This effect continues to be denied by mainstream physics, despite its detection involving at least 8 experiments extending over more than 100 years.  What this effect shows is that reality involves a dynamical 3-space, as Lorentz suggested, and not a spacetime as Einstein suggested. Nevertheless, as discussed in \cite{Coax}, one can arrive at the  spacetime as a well-defined mathematical construct, but which has no ontological significance.  This means that the special  relativity effects are caused by the actual absolute motion of systems through the 3-space as Lorentz long ago suggested.    It also means that this 3-space is a dynamical system and the internal dynamics for this 3-space have already been determined \cite{Book}, and which has lead to a new explanation for gravity, namely that it is caused by the refraction of either EM waves or quantum matter waves by the time dependence and inhomogeneities of the flow of the substructure of this 3-space. As discussed in  \cite{Book,Coax} many of these absolute motion experiments revealed fluctuations or turbulence in the velocity ${\bf v}$, and these correspond to the gravitational waves. These wave effects occur in $v$ at the 20\% level, so even they could be detected in a modern  mechanical light chopper apparatus, although the new optical fiber  technique is even simpler. 

This research is supported by an Australian Research Council Discovery Grant 2005-2006:  {\it Development and Study of a New Theory of Gravity. }

\end{document}